\documentclass[%
 aip,
 amsmath,amssymb,
 reprint,%
]{revtex4-1}

\usepackage{amsmath,amssymb}
\usepackage{graphicx}
\usepackage{xspace,soul,color}

\usepackage{xcolor}

\definecolor{myred}{rgb}{1,0.8,0.8}

\definecolor{mycyan}{rgb}{0.5,0.92,1.0}

\definecolor{mycyan}{rgb}{0.5,0.92,1.0}
\definecolor{mygreen}{rgb}{0.56,0.93,0.56}

\definecolor{myhl}{rgb}{1.0,0.98,0.56}
\sethlcolor{myhl}

\usepackage{graphicx}% Include figure files
\usepackage{dcolumn}% Align table columns on decimal point
\usepackage{bm}% bold math
\usepackage[utf8]{inputenc}
\usepackage[T1]{fontenc}
\usepackage{mathptmx}
\usepackage{etoolbox}
\usepackage{soul} % 导入 soul 包
\usepackage{color, xcolor} % 颜色包，color 必须导入，xcolor 建议导入

\soulregister{\cite}7 % 注册\cite命令
\soulregister{\citep}7 % 注册\citep命令
\soulregister{\citet}7 % 注册\citet命令
\soulregister{\ref}7 % 注册\ref命令

\makeatletter
\def\@email#1#2{%
 \endgroup
 \patchcmd{\titleblock@produce}
  {\frontmatter@RRAPformat}
  {\frontmatter@RRAPformat{\produce@RRAP{*#1\href{mailto:#2}{#2}}}\frontmatter@RRAPformat}
  {}{}
}%
\makeatother
\begin{document}

\preprint{AIP/123-QED}

\title{A spring pair method of finding saddle points using the minimum energy path as a compass}
% Force line breaks with \\
 \author{Gang Cui} \affiliation{ 
 Hunan Key Laboratory for Computation and Simulation in Science and Engineering, Key Laboratory of Intelligent Computing and Information Processing of Ministry of Education, School of Mathematics and Computational Science, Xiangtan University, Xiangtan, Hunan, China, 411105.
 }
 \author{Kai Jiang $^*$}
 \email{kaijiang@xtu.edu.cn}
%  \email{Second.Author@institution.edu.}
 \affiliation{
 Hunan Key Laboratory for Computation and Simulation in Science and Engineering, Key Laboratory of Intelligent Computing and Information Processing of Ministry of Education, School of Mathematics and Computational Science, Xiangtan University, Xiangtan, Hunan, China, 411105.
 }

%\author{Gang Cui and Kai Jiang}\email{kaijiang@xtu.edu.cn}
%\affiliation{Hunan Key Laboratory for Computation and Simulation in Science and Engineering, Key Laboratory of Intelligent Computing and Information Processing of Ministry of Education, School of Mathematics and Computational Science, Xiangtan University, Xiangtan, Hunan, China, 411105.
%}%

\date{\today}% It is always \today, today,

\begin{abstract}
Finding index-1 saddle points is crucial for understanding phase transitions. In this work, we propose a simple yet efficient approach, the spring pair method (SPM), to accurately locate saddle points. Without requiring Hessian information, SPM evolves a single pair of spring-coupled particles on the potential energy surface. By cleverly designing complementary drifting and climbing dynamics based on gradient decomposition, the spring pair converges onto the minimum energy path (MEP) and spontaneously aligns its orientation with the MEP tangent, providing a reliable ascent direction for efficient convergence to saddle points. 
SPM fundamentally differs from traditional surface walking methods, which rely on the eigenvectors of Hessian that may deviate from the MEP tangent, potentially leading to convergence failure or undesired saddle points.
The efficiency of SPM for finding  saddle points is verified by ample examples, including high-dimensional Lennard-Jones cluster rearrangement and the Landau energy functional involving quasicrystal phase transitions.
\end{abstract} 

\maketitle

\section{\label{sec:level1}Introduction}\protect
Locating index-1 saddle points on potential energy surfaces (PES) is crucial for understanding and predicting various phenomena in physics, chemistry, and materials science~\cite{baker1986algorithm,wales2004energy,caspersen2005Finding,Heyden2005Efficient,zhang2007Morphology,cheng2010nucleation,zhang2012noise,amit2014Microscopic,yin2021transition}. 
A typical example is nucleation, where the critical nucleus corresponds to a saddle point on the PES. Critical nuclei emerge transiently in physical experiments and are difficult to observe directly, therefore numerical techniques for computing saddle points become indispensable for investigating such phenomena in phase transitions.

Existing methods for computing saddle points can be classified into two main categories: path-finding methods and surface-walking methods.
Path finding methods, such as the nudged elastic band (NEB)~\cite{henkelman2000improved,Berne1998Classical}, string~\cite{carilli2015truncation,e2002string,samanta2013optimization}, and equal-bond-length~\cite{cui2023finding}, require an initial path connecting the initial and final states. They evolve the path towards the minimum energy path (MEP) by minimizing the energy along the directions perpendicular to the path while maintaining equal spacing between images. The saddle point is then identified as the energy maximum along the MEP. However, these methods may struggle to accurately locate the saddle point when the energy barrier is narrow compared to the total path length, as too few images land near the saddle point~\cite{henkelman2000climbing}.

The climbing image methods, such as the climbing image NEB (CI-NEB)\cite{henkelman2000climbing} and the climbing string method (CSM)\cite{Ren2013climbing}, improve the accuracy by utilizing the MEP tangent as an ascent direction. 
The MEP tangent naturally points towards the saddle point, providing an ideal ascent direction. In climbing image methods, a selected image is driven towards the saddle point by inverting the tangential force while preserving the perpendicular forces. However, climbing image methods still need to optimize the entire path to reach the MEP, which can be computationally expensive, especially for high-dimensional systems. If the primary goal is to locate the saddle point, evolving the entire path may not be necessary.

Surface walking methods~\cite{GM1971minimization,e2011gentlest,miron2001step,cui2023efficient} evolve a single state starting from a known minimum, rather than evolving the entire path. Many surface walking methods, such as the gentlest ascent dynamics (GAD)~\cite{e2011gentlest} and the dimer method~\cite{henkelman1999dimer}, belong to min-mode methods, which rely on the eigenvector corresponding to the smallest eigenvalue of Hessian (i.e., the lowest curvature mode) to determine the ascent direction. However, computing the Hessian can be expensive for large systems. The dimer method can avoid explicit Hessian calculation by rotating the dimer to find the lowest curvature mode. However, since the lowest curvature mode is a local information, it may deviate from the MEP tangent, potentially causing the min-mode methods to diverge from the MEP~\cite{henkelman1999dimer,Josep2013Locating,Bofill2015variational} and fail to converge to the desired saddle point. 
Since the MEP is the most probable path connecting the minimum and the saddle point, the MEP tangent naturally provides an ascent direction towards the saddle point. This motivates us to develop a surface walking method that ensures the climbing direction aligns with the MEP tangent.

Inspired by path-finding and surface-walking methods, we propose the spring pair method (SPM) for efficiently locating saddle points by utilizing the MEP tangent as an ascent direction without evolving the entire path. Specifically, SPM employs two carefully designed dynamics, drifting and climbing, to evolve a pair of particles connected by a spring for locating saddle points on the PES.
The drifting dynamics uses the gradient component perpendicular to the spring orientation to steer the spring pair towards the MEP while maintaining an appropriate distance between the particles through the spring force. This process naturally aligns the spring orientation with the MEP tangent, providing a reliable ascent direction for the subsequent climbing dynamics.
The climbing dynamics uses the gradient component parallel to the spring orientation as the ascending force to propel the spring pair towards the saddle point along the MEP tangent direction. 
By leveraging the MEP tangent as a reliable ascent direction and evolving only a spring pair, SPM ensures efficient convergence to the saddle point with economical computational costs.

Furthermore, we demonstrate  the power of SPM by studying two-dimensional functions and high-dimensional problems, including the Lennard-Jones (LJ) cluster rearrangement and phase transitions of quasicrystals based on the Lifshitz-Petrich (LP) energy functional. 
The two-dimensional functions validate the basic operating principles and advantages over min-mode methods.
Furthermore, the precise location of the transition state of the LJ cluster and the critical nuclei of quasicrystals prove that SPM is an efficient tool for exploring complex, high-dimensional PES.

\section{Spring Pair Method}
We present the implementation of SPM to locate  saddle points
from a local minimum on a PES $E(\mathbf{r})$. This is achieved by constructing a simple spring pair system and fully utilizing gradient information and intrinsic nature of MEP, without requiring Hessian calculations.

\begin{figure}
        \centering	
		\includegraphics[width=1.0\columnwidth]{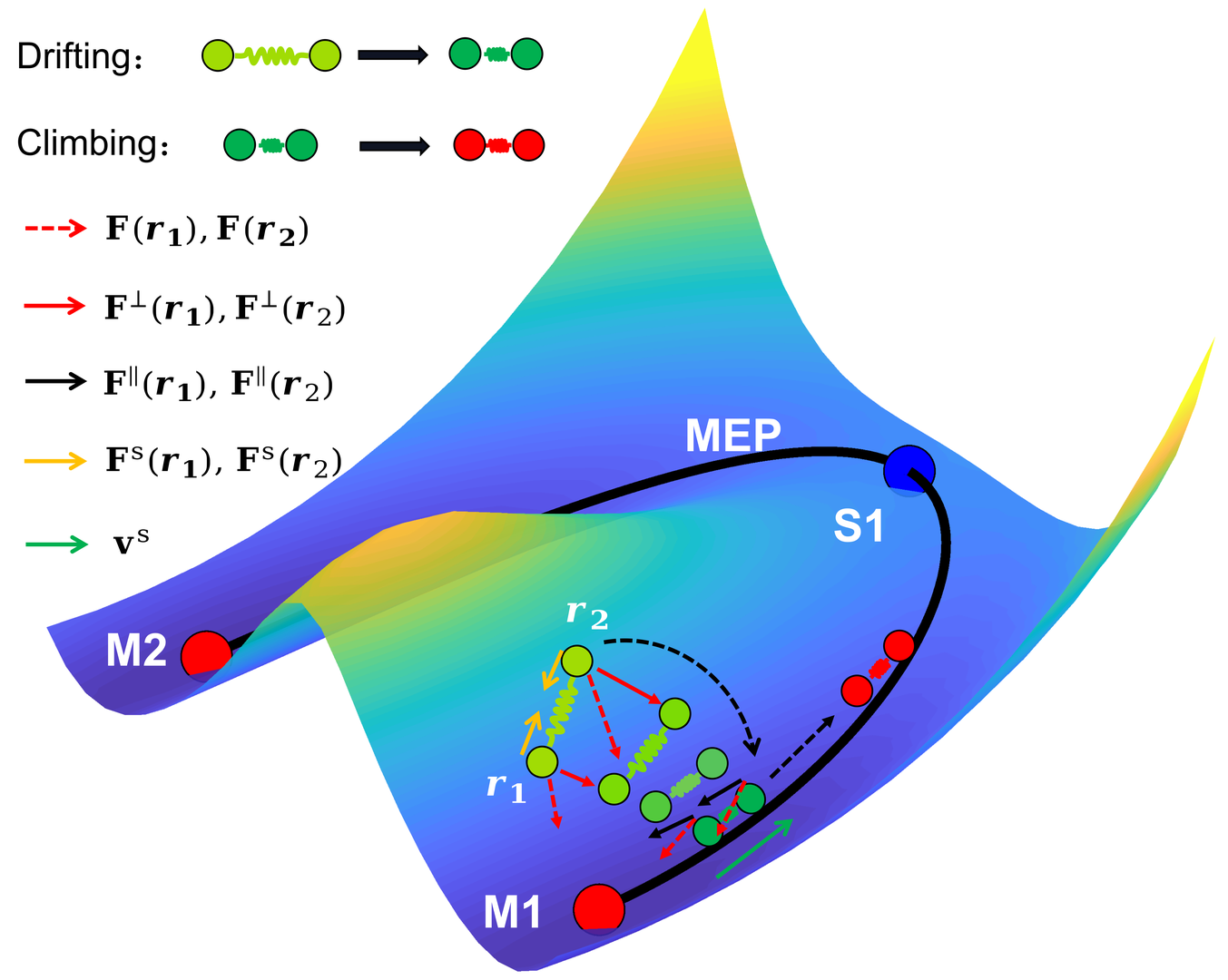}
        \caption{Schematic illustration of the designed drifting and climbing dynamics. The black curve represents the MEP connecting two minima, M1 and M2, through a saddle point S1. The black dotted curved and straight arrows indicate the directions of drifting and climbing, respectively. During the drifting process, the perpendicular negative gradient component $\mathbf{F}^\perp(\mathbf{r}_i)$ (red arrow) drives the spring pair (light green to green) towards the MEP. Concurrently, the spring force $\mathbf{F}^s(\mathbf{r}_i)$ (yellow arrow) spontaneously adjusts the spring length. Consequently, the orientation $\mathbf{v}^s$ (green arrow) of the spring pair (green) on the MEP becomes the MEP tangent. In the climbing process, the spring pair (green) moves along the MEP tangent, driven by the parallel gradient component $-\mathbf{F}^\parallel(\mathbf{r}_i)$ (opposite black arrow). After one climbing step, the spring pair (red) reaches a higher energy position closer to the saddle point S1.}\label{FIG1}
\end{figure}

% {\it Construct Spring Pair.--}
\subsection{\label{sec:level2}Constructing the spring pair}
As illustrated in Fig.\,\ref{FIG1}, the spring pair system contains two particles $\mathbf{r_1}, \mathbf{r_2}$ connected by a spring with natural length $l_s$. 
The current spring length is denoted as $d_s = |\mathbf{r}_1 - \mathbf{r}_2|$, and the spring force on each particle is given by:
\begin{equation}
\begin{aligned}
\mathbf{F}^s(\mathbf{r_1}) &= (d_s - l_s)(\mathbf{r_2} - \mathbf{r_1}), \\
\mathbf{F}^s(\mathbf{r_2}) &= (d_s - l_s)(\mathbf{r_1} - \mathbf{r_2}).
\end{aligned}
\end{equation}
When $d_s < l_s$, the spring is compressed and $\mathbf{F}^s$ separates $\mathbf{r_1}$ and $\mathbf{r_2}$. While $d_s > l_s$, the spring is stretched and $\mathbf{F}^s$ pulls $\mathbf{r_1}$ and $\mathbf{r_2}$ closer.
The natural length $l_s$ of the spring is a hyper-parameter that needs to be properly chosen to ensure that the spring direction reliably represents the MEP tangent during the drifting process.
Typically, $l_s$ is can be sufficiently small, related to the system's energy or  gradient scale. 
% In practice, an appropriate value for $l_s$ can be easily determined through a few numerical experiments, as it usually only requires adjustments on an exponential scale, such as $10^{-3}$ or $10^{-4}$, rather than precise adjustment.
The spring force, determined by $l_s$, effectively adjusts the inter-particle distance, preventing the particles from separating too far or overlapping each other. This allows the spring orientation to accurately capture the MEP tangent direction in the subsequent drifting  dynamics.

Let $\mathbf{F}(\mathbf{r}) = -\nabla E(\mathbf{r})$ be the negative gradient, and $\mathbf{v}^{s} = (\mathbf{r}_2 - \mathbf{r}_1)/(|\mathbf{r}_2 - \mathbf{r}_1|)$ be the spring direction. We decompose $\mathbf{F}(\mathbf{r}_i)$ into parallel and perpendicular components with respect to $\mathbf{v}^{s}$:
\begin{equation}
\begin{aligned}
\mathbf{F}^{\parallel}(\mathbf{r_i}) &= (\mathbf{F}(\mathbf{r_i}) \cdot \mathbf{v}^{s})\mathbf{v}^{s}, \\
\mathbf{F}^{\perp}(\mathbf{r_i}) &= \mathbf{F}(\mathbf{r_i}) - \mathbf{F}^{\parallel}(\mathbf{r_i}).
\end{aligned}
\end{equation}

The MEP is the most probable transition path connecting the local minima and the saddle points, along which the energy is minimized in all directions in the variable space except for the path tangent. Thus, the MEP can be used as a global compass to guide the dynamics of the SPM, and the MEP tangent serves as a natural ascent direction for escaping the minimum towards the saddle points. The drifting dynamic guides the spring pair onto the MEP and aligns the spring orientation with the MEP tangent. The climbing dynamic then drives the spring pair along the MEP tangent toward higher energy. By alternating between drifting and climbing steps, the spring pair can move along the MEP and eventually converge to the saddle points.

% {\it Drifting.--}
\subsection{Drifting}
As illustrated in Fig.\,\ref{FIG1}, the perpendicular component of the gradient $\mathbf{F}^{\perp}(\mathbf{r}_i)$ steers the spring pair towards the MEP, similar to the path-finding methods. However, unlike path-finding methods that aim to obtain the entire MEP, SPM only requires the spring pair to fall onto the MEP. 
Concurrently, the spring force $\mathbf{F}^s(\mathbf{r}_i)$ automatically adjusts the inter-particle distance $d_s$, preventing them from separating too far or overlapping each other.
The dynamics of drifting is
\begin{equation}\label{dynamic_drift}
\dot{\mathbf{r}}_i = \alpha_1 \mathbf{F}^{\perp}(\mathbf{r}_i) + \alpha_2 \mathbf{F}^s(\mathbf{r}_i),
\end{equation}
where $\alpha_1,\alpha_2$ are positive relaxation constants. 
When $\max{\lbrace |\mathbf{F}^{\perp}(\mathbf{r_1})|, |\mathbf{F}^{\perp}(\mathbf{r_2})| \rbrace} < \epsilon_1$, $\epsilon_1$ is a small positive constant, the spring pair $\mathbf{r}_1, \mathbf{r}_2$ is considered to be on the MEP, with the spring orientation $\mathbf{v}^s$ aligned with the MEP tangent.

The drifting process in SPM is fundamentally different from the rotation step in the dimer method~\cite{henkelman1999dimer}. The dimer rotation only changes the orientation of the dimer while keeping its center position fixed, aiming to find the lowest curvature mode as the ascent direction. In contrast, the SPM drifting process  adjusts both the orientation and position of the spring pair to directly obtain the MEP tangent as the ascent direction. By using the perpendicular component of the gradient and the spring force, the drifting process steers the spring pair onto the MEP, naturally aligning the spring orientation with the MEP tangent. 
% This key difference enables SPM to acquire a reliable ascent direction that inherently points towards the saddle point.
% This key difference enables the SPM drifting to acquire a reliable ascent direction for the subsequent climbing process, since the MEP tangent naturally points towards the saddle point.

% {\it Climbing.--}
\subsection{Climbing}
Once the spring pair has drifted onto the MEP, the spring direction $\mathbf{v}^s$ aligns with the MEP tangent, providing a natural ascent direction towards the saddle point.
The climbing process is driven by $-\mathbf{F}^{\parallel}(\mathbf{r}_i)$, which is the projection of the gradient onto $\mathbf{v}^s$, as illustrated in Fig.\,\ref{FIG1}. The climbing dynamics is 
\begin{equation}\label{dynamic_climb}
\dot{\mathbf{r}}_i = -\alpha_3 \mathbf{F}^{\parallel}(\mathbf{r}_i),
\end{equation}
where $\alpha_3$ is a relaxation parameter.

% {\it Drift-Climb-Cycle.--}
\subsection{Drift-Climb-Cycle}
After each climbing step, the spring pair moves to a higher energy position along MEP tangent but may slightly deviate from the MEP. Drifting is then performed to bring the spring pair back to the MEP.
By utilizing the MEP as a global compass, the SPM alternates between drifting and climbing dynamics, ensuring that the spring pair moves consistently towards the saddle point while maintaining proximity to the MEP.
The drift-climb-cycle continues until the following condition is satisfied
\begin{equation}\label{e_1def}
e_1 = \min{\lbrace\lvert \mathbf{F}(\mathbf{r_1}) \rvert, \lvert \mathbf{F}(\mathbf{r_2}) \rvert \rbrace} < \epsilon_2,
\end{equation}
where $\epsilon_2$ is a small threshold. When $e_1 < \epsilon_2$, a particle of the spring pair converges to the  saddle point along the MEP, with $\mathbf{v}^{s}$ aligning with the unstable mode at the saddle point. Throughout the evolution process, the MEP serves as a global compass, guiding the spring pair towards the saddle point and ensuring efficient convergence by consistently providing a reliable ascent direction. 

Since the ultimate goal of SPM is to accurately locate the saddle points, the termination condition for the drifting process can be relaxed, and the maximum number of drifting iterations can be adjusted flexibly. This allows the spring pair to efficiently converge to the saddle point without strictly following to the MEP at every step of the evolution process, thereby reducing computational cost.
Once the saddle point and its corresponding unstable mode have been identified, an accurate MEP connecting the minima can be easily obtained by slightly perturbing the system along the unstable mode at the saddle point and subsequently performing gradient descent towards the minimum.
% {\it Discrete dynamics.--}
\subsection{\label{sec:level2}Discrete Dynamics}
The dynamics described in Equations \eqref{dynamic_drift} and \eqref{dynamic_climb} can be integrated in time using any suitable discrete scheme, such as the forward Euler method, the Runge-Kutta method, or the backward differentiation formula. Here the forward Euler scheme is utilized to illustrate the numerical implementation procedure.

Let $\mathbf{r}^{(n)}_i$, $i=1,2$, denote the positions of the particles in the spring pair after $n$ drifting iterations. A single drifting update with timestep $\Delta t$ is given by
\begin{equation}
\mathbf{r}^{(n+1)}_i = \mathbf{r}^{(n)}_i + \Delta t \left[ \alpha_1 \mathbf{F}^{\perp}(\mathbf{r}^{(n)}_i) + \alpha_2 \mathbf{F}^s(\mathbf{r}^{(n)}_i) \right].
\end{equation}
For simplicity, we rewrite $\alpha_1 =  \alpha_1 \Delta t$, $\alpha_2 =  \alpha_2 \Delta t$ as scaled relaxation parameters. The iterative format for drifting is given by
\begin{equation}\label{Euler_dynamic_drift}
\mathbf{r}^{(n+1)}_i = \mathbf{r}^{(n)}_i +  \alpha_1 \mathbf{F}^{\perp}(\mathbf{r}^{(n)}_i) + \alpha_2 \mathbf{F}^s(\mathbf{r}^{(n)}_i) .
\end{equation}
The drifting dynamics is terminated when $\max{\lbrace |\mathbf{F}^{\perp}(\mathbf{r^{(n)}_1})|, |\mathbf{F}^{\perp}(\mathbf{r^{(n)}_2})| \rbrace} < \epsilon_1$,  or the maximum number of steps $n^{(d)}$ is reached. 

Let $\mathbf{r}^{(*)}_i,i=1,2$, represent the positions of the spring pair after the drifting process in the $m$-th drift-climb-cycle. A single climbing update is then given by

\begin{equation}\label{Euler_dynamic_climb}
\mathbf{r}^{(m)}_i = \mathbf{r}^{(*)}_i  -\alpha_3 \mathbf{F}^{\parallel}(\mathbf{r}^{(*)}_i),
\end{equation}
where $\alpha_3 = \alpha_3\Delta t$ is the scaled relaxation parameter. 

% \subsection{Initial Spring Pair Setting}

%\subsubsection{\label{sec:level3}Third-level heading: Citations and Footnotes}

\section{Numerical Examples}
In this section, we demonstrate the effectiveness of the SPM through a series of examples, ranging from simple two-dimensional test functions to high-dimensional problems in  physical systems. We begin by analyzing SPM's behavior on 2D examples, which provide a visually intuitive way to understand its operating mechanisms and highlight its advantages over existing methods. We then showcase SPM's applicability to more complex systems, including the LJ cluster and the LP energy functional for quasicrystal phase transitions. These examples demonstrate SPM's ability to efficiently locate saddle points on high-dimensional PES.

% {\it Two-dimensional function.--}
\subsection{Two-dimensional functions}
\begin{figure}
        \centering	
		\includegraphics[width=0.48\columnwidth]{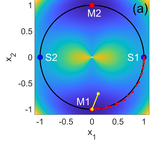}~~~
		\includegraphics[width=0.48\columnwidth]{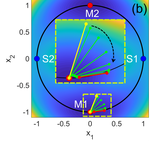}\\~~~\hspace{0.00000000001cm} 
		\includegraphics[width=0.48\columnwidth]{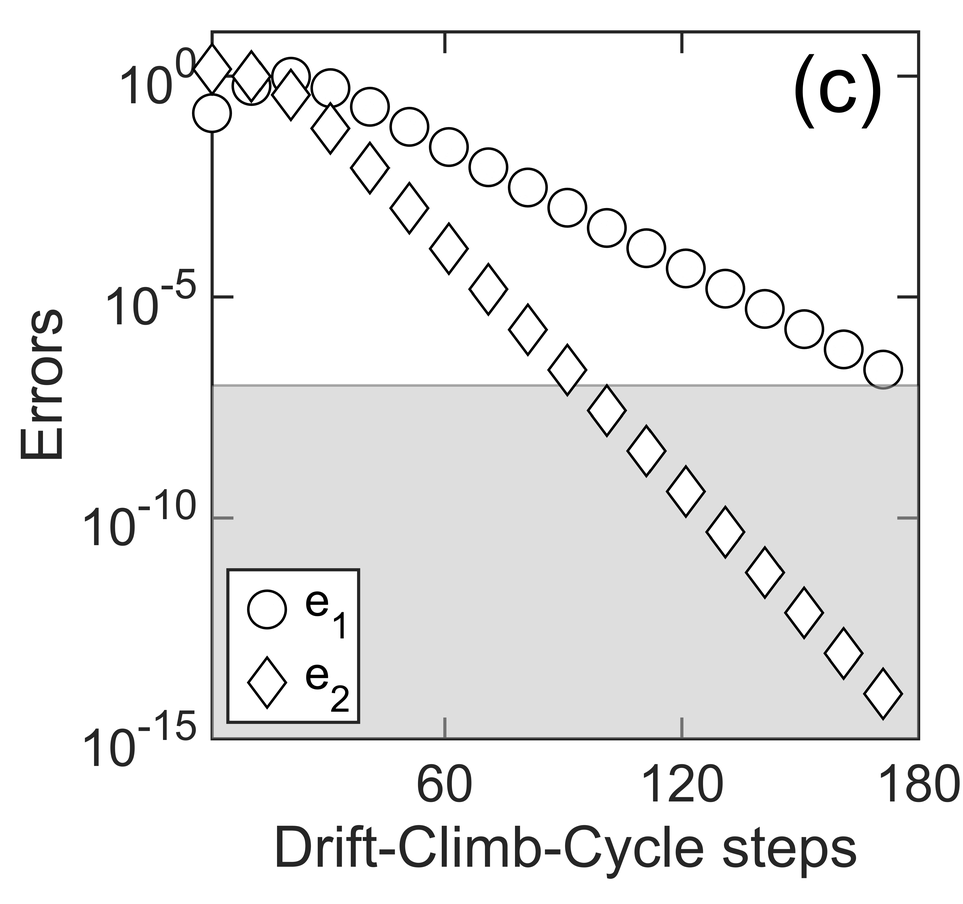}~~~
        \includegraphics[width=0.48\columnwidth]{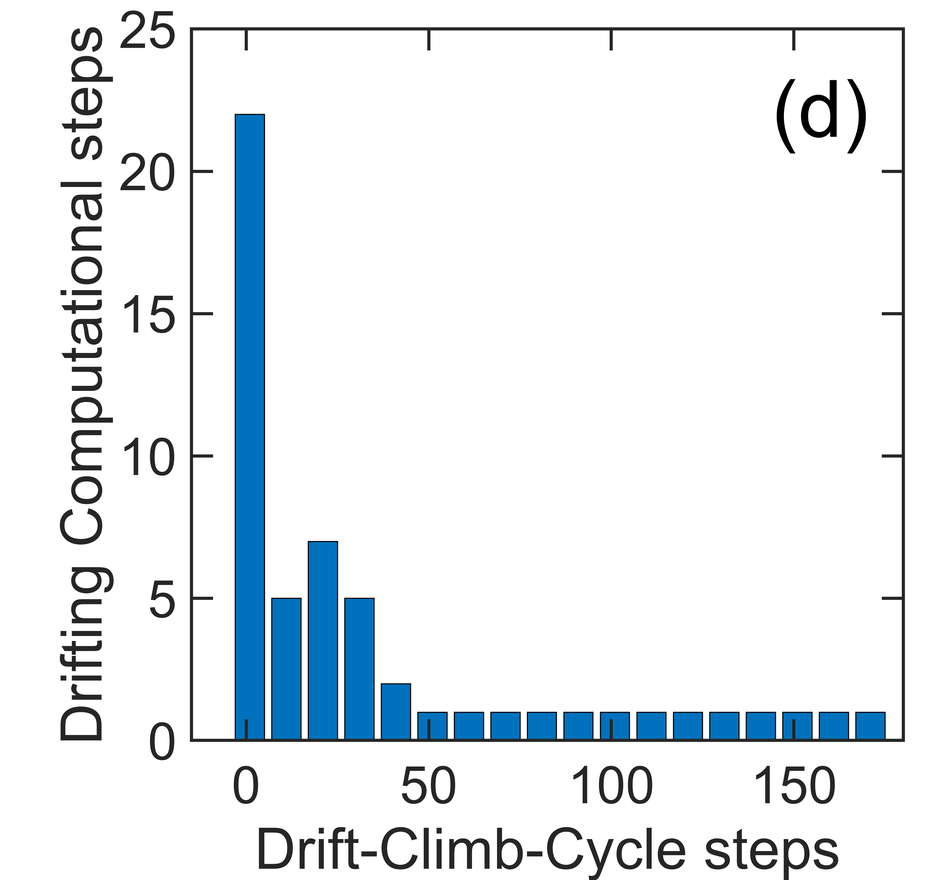}\\
        \caption{Visualised analysis of SPM on the test function $V_1(\mathbf{x}_1, \mathbf{x}_2)$. (a, b) Local minima (M1, M2, red points) and saddle points (S1, S2, blue points) are annotated, and the exact MEP is depicted as the black curve. Yellow spheres connected by solid lines indicate the initial spring pair. (a) The evolution process of the spring pair from M1 to S1 along the MEP. (b) The drifting process drives the spring pair onto the MEP, exemplified by the initial yellow pair evolving to the red pair on the MEP. (c) Convergence of errors $e_1$ and $e_2$ versus drift-climb-cycle steps. The tolerance level $\epsilon_2$ in \eqref{e_1def} is represented by the line dividing the white and grey regions. (d) Number of drifting steps taken within each drift-climb-cycle, showing fewer drifting steps needed as the spring pair approaches the saddle point. 
          }\label{FIG2}
\end{figure}

We first apply SPM to find saddle points on a two-dimensional function
\begin{equation}\label{V_1}
V_{1}(\mathbf{x}_1, \mathbf{x}_2) = (1-\mathbf{x}_1^2-\mathbf{x}_2^2)^2 + \frac{\mathbf{x}_1^2}{(\mathbf{x}_1^2+\mathbf{x}_2^2)},
\end{equation}
which is a proper case that also been used in the string method~\cite{e2002string}.
As depicted in Fig.\,\ref{FIG2}\,(a,b), $V_1(\mathbf{x}_1, \mathbf{x}_2)$ has two minima M1$ (0, -1)$ and M2$(0, -1)$, and two saddle points S1$(1, 0)$ and S2$(-1, 0)$. The MEP connecting the minima through the saddle points follows the circle defined by $\mathbf{x}_1^2+\mathbf{x}_2^2=1$, the black curves in Fig.\,\ref{FIG2}\,(a,b). 
Furthermore, the origin $(0,0)$ is a singularity at which the function approaches the infinity.

\subsubsection{Performance of SPM}
The goal is to find the saddle point S1 starting from the minimum M1. The initial spring particles are $\mathbf{r}^{0}_1 = (0,-1)$ at M1 and $\mathbf{r}^{0}_2 = \mathbf{r}^{0}_1 + \epsilon \mathbf{v}^{*}$ is a perturbation of $\mathbf{r}^{0}_1$, shown as two yellow spheres in Fig. \,\ref{FIG2}\,(a,b). Here we set the perturbation direction $\mathbf{v}^{*} = (0.4, 1)$, the perturbation size $\epsilon = 0.3$. The parameters used in SPM are $l_s=10^{-2}$, $\alpha_1=\alpha_3=5\times 10^{-2}$, $\alpha_2=2.5\times 10^{-1}$. Convergence criterion is $\epsilon_2=10^{-7}$, drifting criterion is $\epsilon_1=10^{-2}$ and maximum iteration in drifting process is $n^{(d)} = 200$. To quantify accuracy of SPM, we define error $e_2 = \min(|\mathbf{r}_1-\text{S1}|,|\mathbf{r}_2-\text{S1}|)$. 

Fig.\,\ref{FIG2}\,(a) shows the evolution process of SPM, where the spring pair searches S1 from M1 along the MEP tangent direction. Fig.\,\ref{FIG2}\,(b) demonstrates the drifting process driving the spring pair onto the MEP, using the initial yellow pair evolving to the red pair on the MEP as an example. The green spring pair in Fig.\,\ref{FIG2}\,(b) shows the drifting process, with the black arrow indicating the drifting direction. The spring pair gradually approaches the MEP driven by $\mathbf{F}^{\perp}(\mathbf{r_i})$ while the spring force automatically adjusts spring length. Eventually, the spring pair falls on the MEP, with the spring direction being the MEP tangent. Fig.\,\ref{FIG2}\,(c) shows the convergence behavior of errors $e_1$ and $e_2$ over the number of drift-climb-cycle. Fig.\,\ref{FIG2}\,(d) illustrates the number of drifting steps for each drift-climb-cycle. Initially, the drifting dynamics requires more steps to bring the initial spring pair to the MEP since the initial state is randomly imposed. Then when the climbing dynamic causes the spring pair to detach from the MEP, the drifting dynamics brings it back again. Finally, as the spring pair approaches the saddle points, 
fewer drifting steps are needed, indicating that the spring pair closely follows the MEP.

\begin{figure}
        \centering	
		\includegraphics[width=0.49\columnwidth]{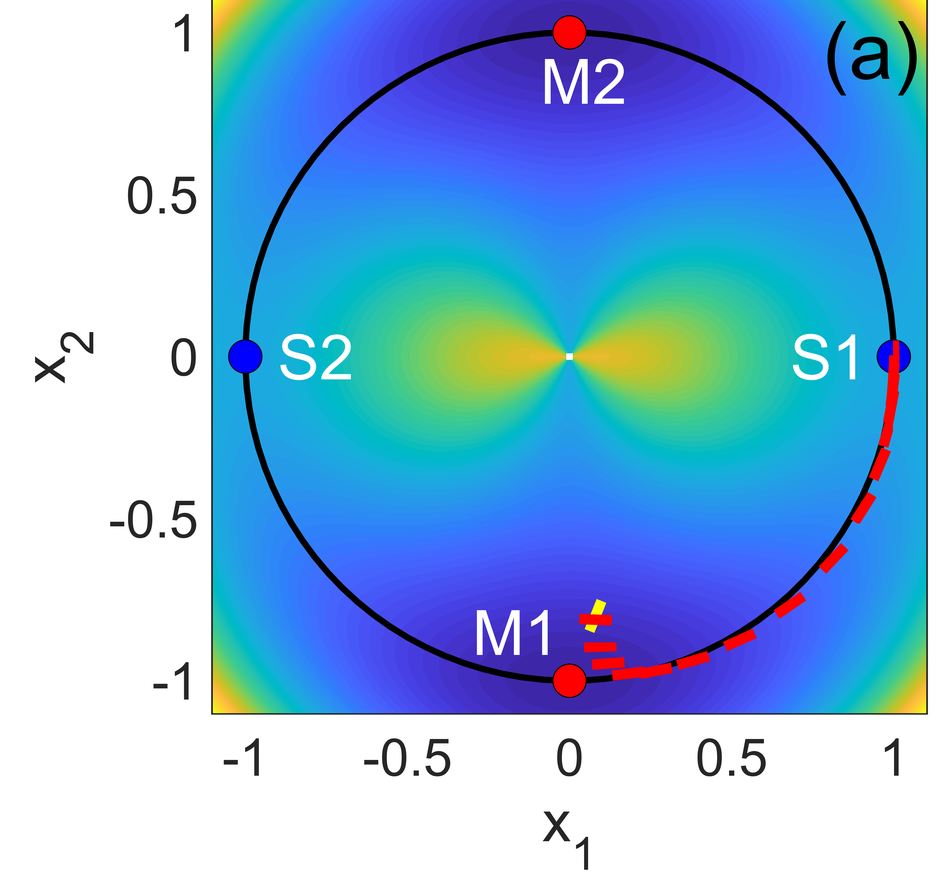}~~~
		\includegraphics[width=0.49\columnwidth]{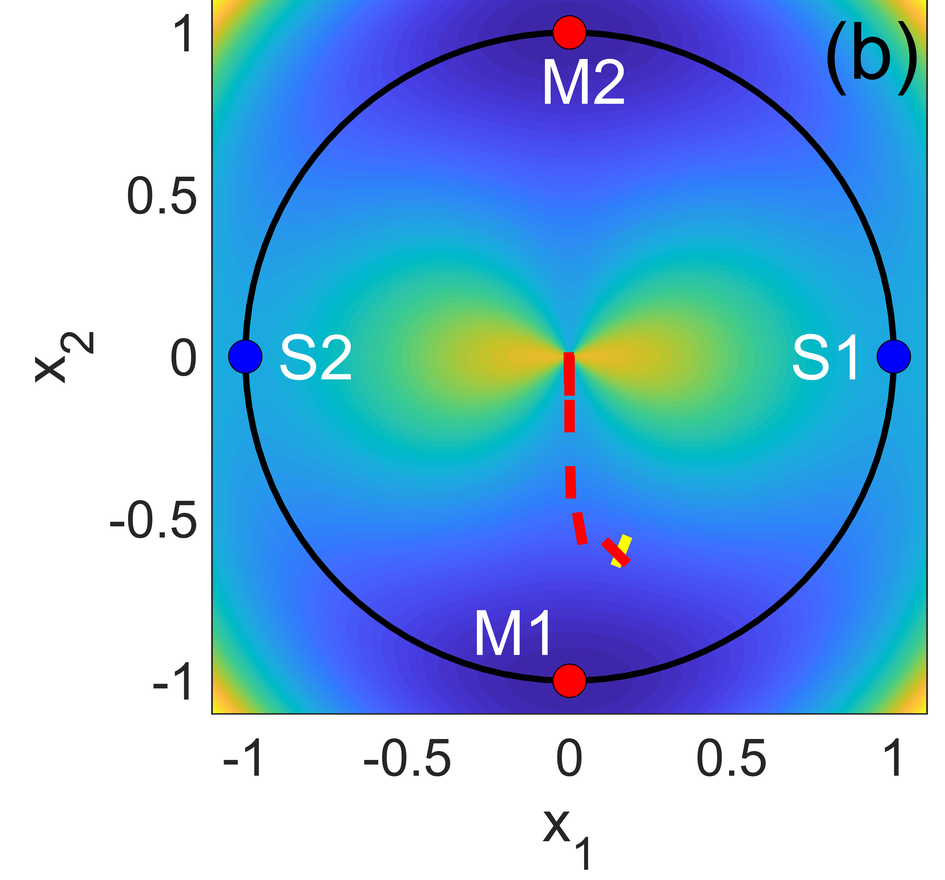}\\
		\includegraphics[width=0.49\columnwidth]{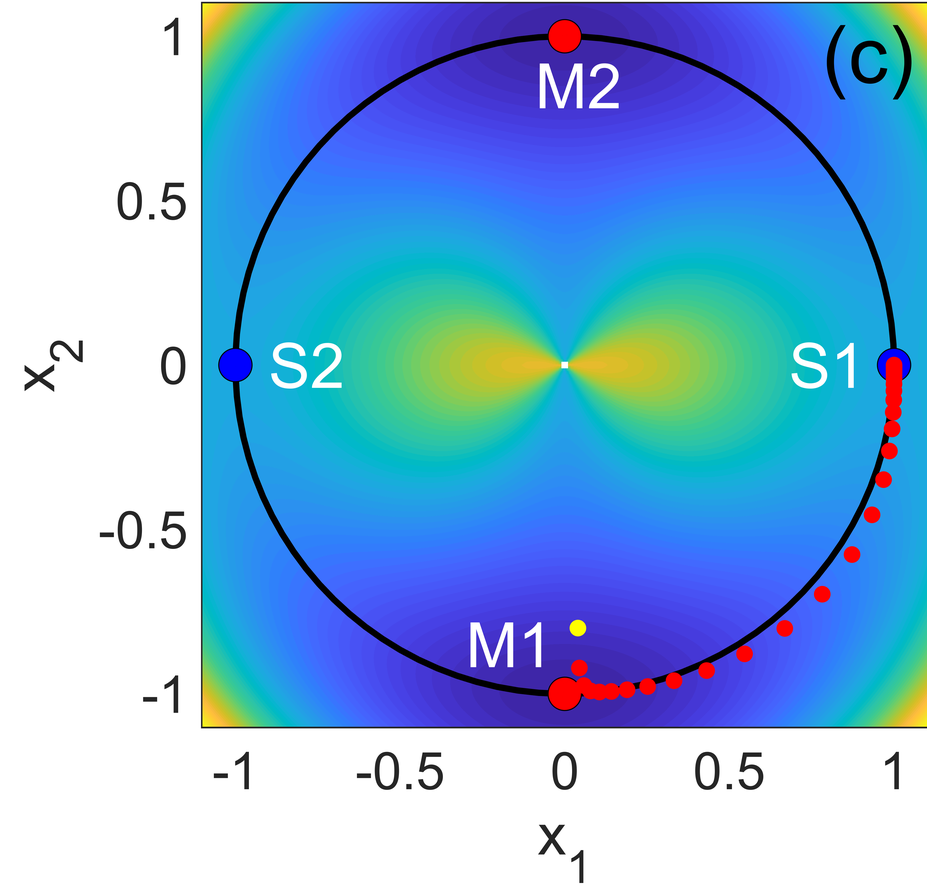}~~~
	    \includegraphics[width=0.49\columnwidth]{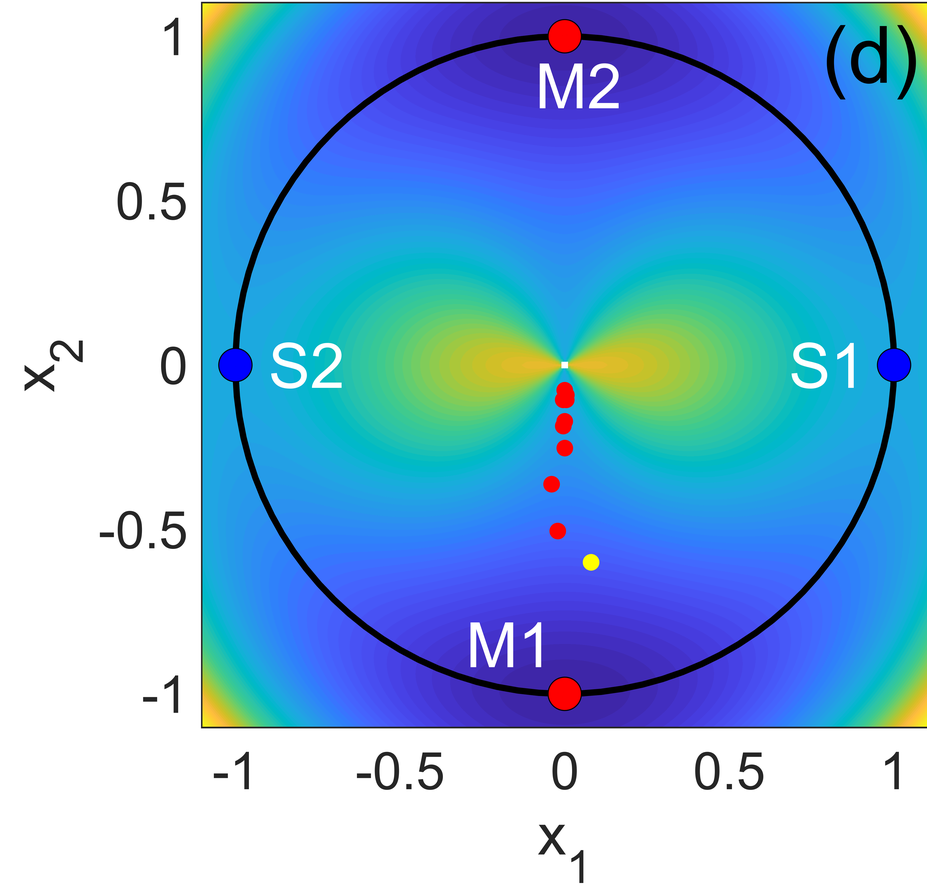}\\
		\includegraphics[width=0.49\columnwidth]{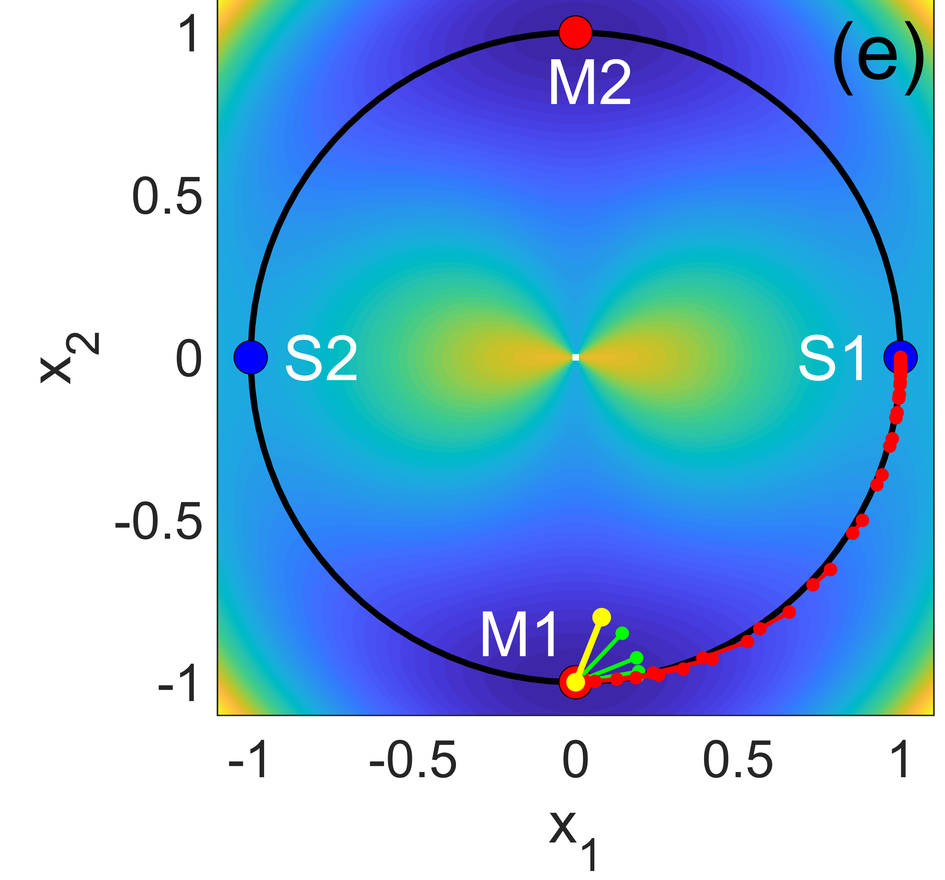}~~~
		\includegraphics[width=0.49\columnwidth]{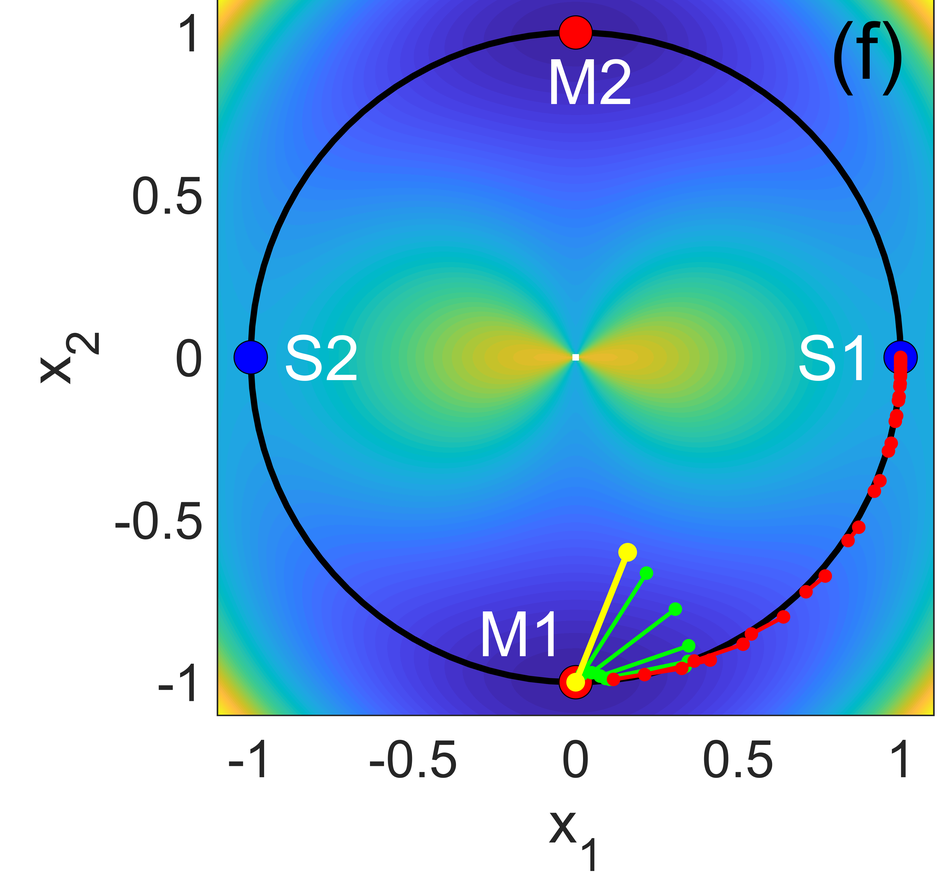}\\
        \caption{Effects of perturbations on dimer (a-b), GAD (c-d) and SPM (e-f) algorithms. The initial configurations are represented by yellow bars (dimer), dots (GAD) and spring pairs (SPM). With a small perturbation from the minimum M1 in the $\mathbf{v}^{*}$ direction, the dimer (a), the GAD (d) and the SPM (e) all converge to the target saddle points S1. However, under large perturbation, dimer (b) and GAD (d) fail to converge and are trapped near singularity (0,0), while SPM (f) still reliably locates S1 following MEP. The green spring pair in (e-f) shows the drifting process from the initial yellow pair, which always drifts towards the MEP, even if the perturbation size is changed.
          }\label{FIG2_min_mode}
\end{figure}

\subsubsection{Differences from the min-mode methods}
Min-mode methods, such as the dimer method and GAD, have been widely used to find the saddle point. However, these methods rely on the lowest curvature mode at each point as the ascending direction, which may deviate from the MEP tangent direction and lead to failure of convergence to the saddle point.

{\it Differences in convergence.--}
To compare the performance of SPM with min-mode methods, we apply the dimer method, GAD, and SPM to the test function $V_1(\mathbf{x}_1,\mathbf{x}_2)$ with different perturbations to some initial states.
As shown in Fig.\ref{FIG2_min_mode}, when starting from a small perturbation to the minimum M1, all three methods (dimer (a), GAD (c), and SPM (e)) successfully converge to the saddle point S1. However, as the perturbation magnitude increases, the dimer method (b) and GAD (d) fail to converge and get trapped in the vicinity of the singular point $(0,0)$. In contrast, SPM (f) maintains its robustness and reliably locates the saddle point S1 by closely following the MEP, irrespective of the perturbation magnitude.

The difference in performance is due to the inherent distinction in how these methods determine the ascent direction.
Min-mode methods rely solely on the local curvature information at each point, which is not necessarily aligned with the MEP tangent. Consequently, their evolution paths can be wrongly attracted to irrelevant regions on the PES, resulting in the failure of convergence, as illustrated in Fig.\ref{FIG2_min_mode}\,(b,d). In contrast, SPM exploits the MEP as an intrinsic and global compass to guide the search dynamics. By following the MEP tangent direction, SPM ensures efficient and reliable convergence to the saddle point connected to the initial state, regardless of the perturbation scale, as demonstrated in Fig.\,\ref{FIG2_min_mode}\,(e,f).

% \subsubsection{\hlcy{The role of spring}}
% \begin{figure}
%         \centering	
% 		\includegraphics[width=0.48\columnwidth]{./fig/SPM_a.png}~~~
% 		\includegraphics[width=0.48\columnwidth]{./fig/SPM_b.png}\\
%         \caption{Stability of SPM to large perturbations. 
%           }\label{SPM_turmoil}
% \end{figure}
\begin{figure}
        \centering	
		\includegraphics[width=0.49\columnwidth]{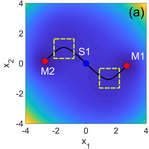}\\
		\includegraphics[width=0.48\columnwidth]{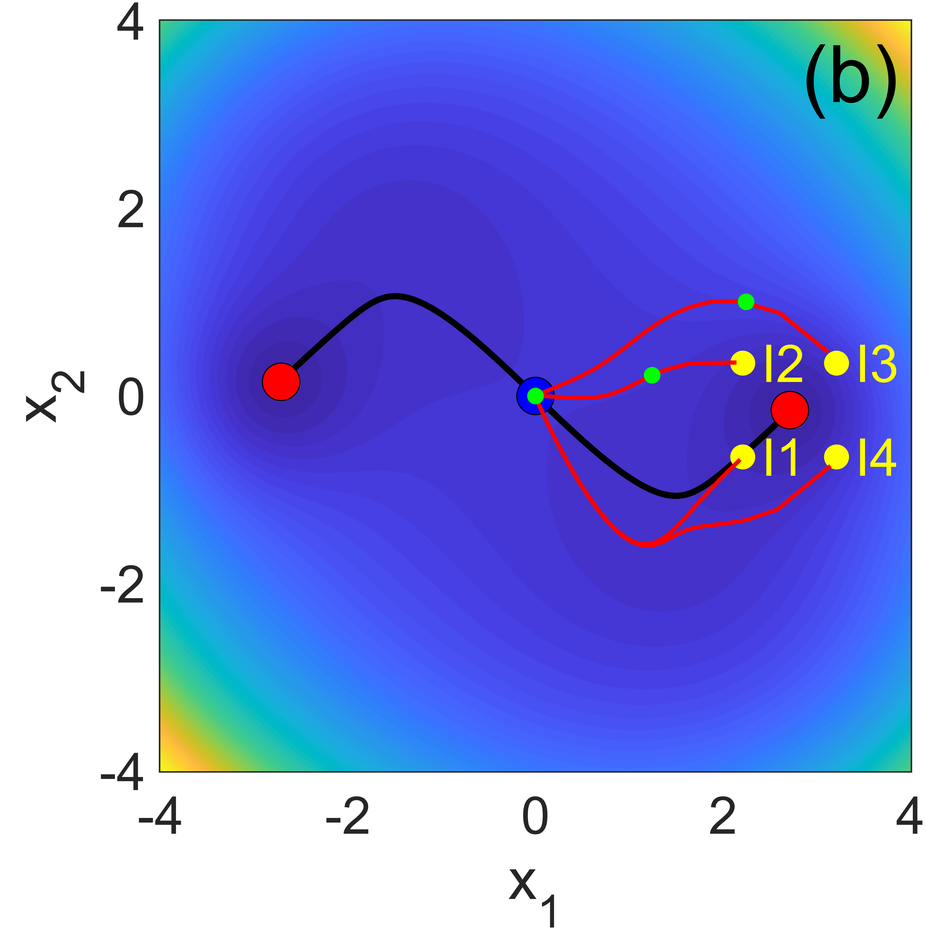}~~~
		\includegraphics[width=0.48\columnwidth]{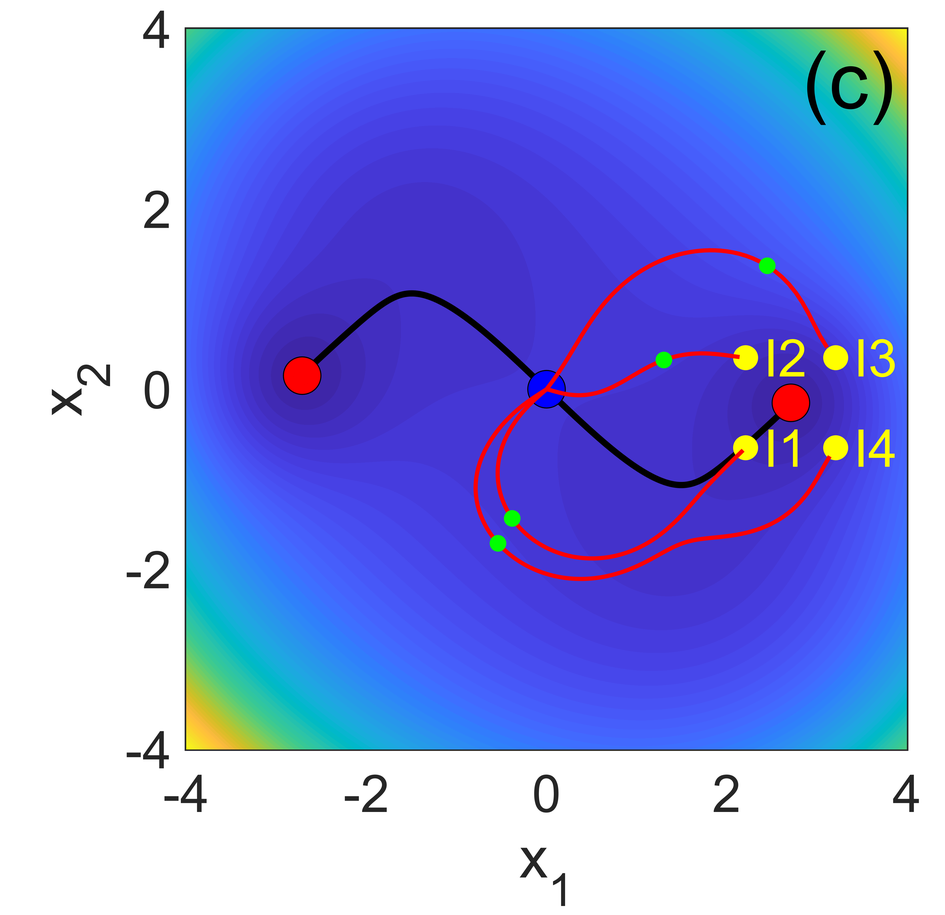}\\
	    \includegraphics[width=0.48\columnwidth]{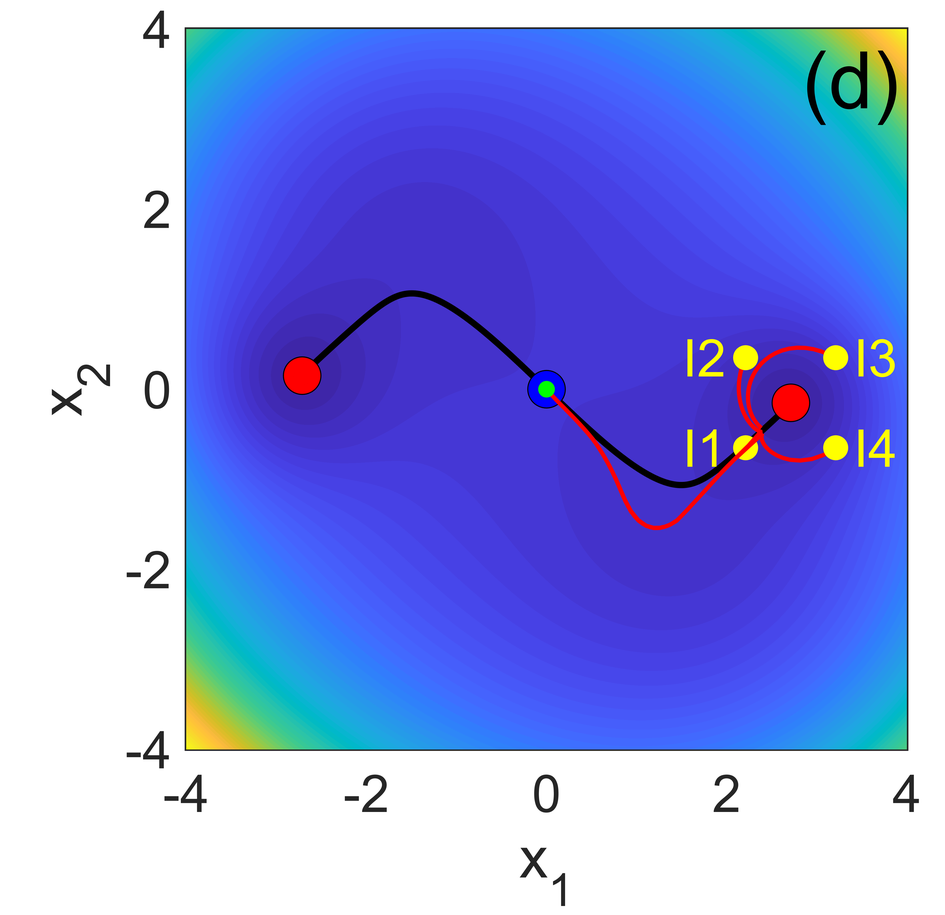}~~~
	    \includegraphics[width=0.48\columnwidth]{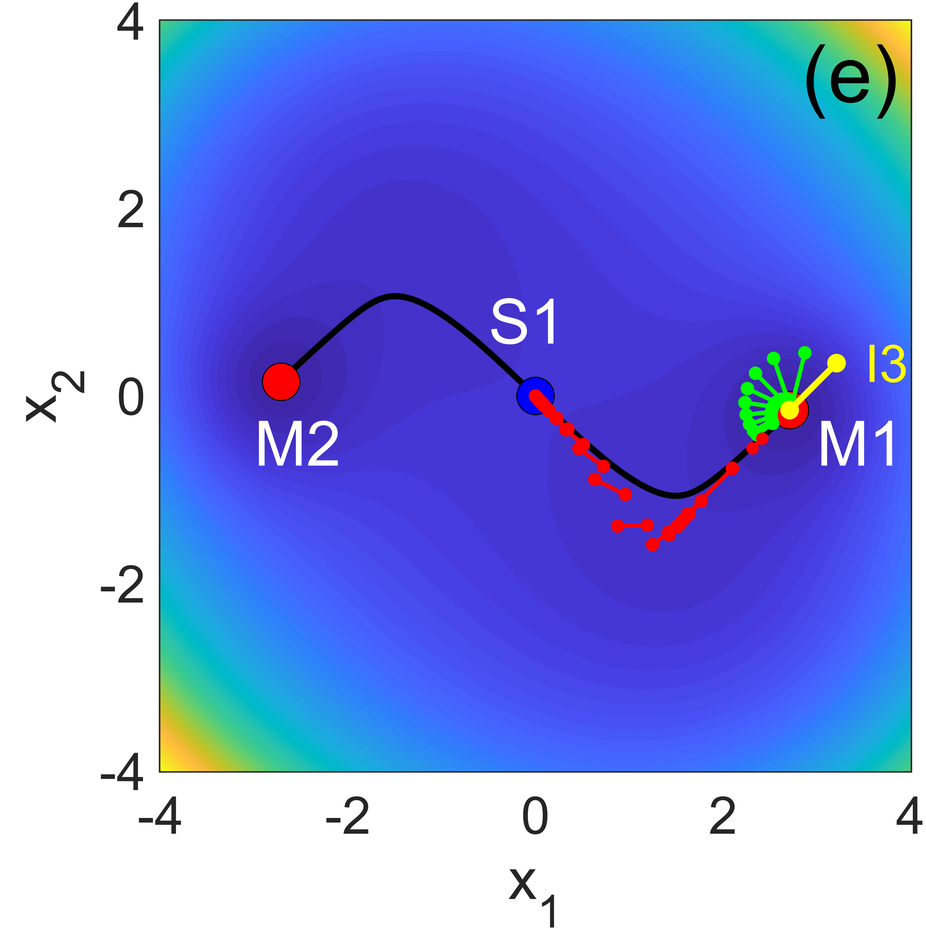}
        \caption{Comparison of the evolution paths of the dimer method, GAD, and SPM on the test function $V_2(\mathbf{x}_1,\mathbf{x}_2)$. (a) The potential energy surface with two minima (M1 and M2, red points), a saddle point (S1, blue point), and the MEP connecting them (black curve). The yellow boxes highlight two sharp right-angle turns where the MEP abruptly changes direction. (b-d) The evolution paths (red curves) of the dimer, GAD, and SPM methods starting from different initial positions (I1, I2, I3, I4, yellow points) obtained by perturbing the minima M1 in the directions of $\pm(1,1)$ and $\pm(-1,1)$. The green points on the curves represent the points with highest energy along the evolution paths. (e) Evolution details of SPM starting from the initial position $I_3$, which is opposite to the MEP direction. The green spring pairs illustrate the drifting dynamics pulling the  initial spring pair back onto the MEP. 
        % (f) The number of drifting steps in each drift-climb-cycle of SPM, showing two peaks corresponding to initial position and sharp right-angle turn. 
          }\label{FIG3_NHQ}
\end{figure}

{\it Differences in evolutionary paths.--}
To further illustrate the differences between SPM and min-mode methods, we consider the modified Neria–Fischer–Karplus PES \cite{Neria1996Simulation,Hirsch2004Reaction}
\begin{eqnarray}\label{V_2}
V_{2}(\mathbf{x}_1, \mathbf{x}_2) &=& 0.06(\mathbf{x}_1^2 + \mathbf{x}_2^2)^2 + \mathbf{x}_1 \mathbf{x}_2 - 9\mathrm{e}^{-(\mathbf{x}_1 -3)^2 - \mathbf{x}_2^2}
\nonumber\\
&-& 9\mathrm{e}^{-(\mathbf{x}_1 +3)^2 - \mathbf{x}_2^2}.
\end{eqnarray}
As shown in Fig.\,\ref{FIG3_NHQ}\,(a), the function $V_2(\mathbf{x}_1,\mathbf{x}_2)$ possesses two minima (M1 and M2) and one saddle point (S1), which are connected by the MEP (depicted by the black curve). Notably, the MEP exhibits two abrupt right-angle turns, where its direction shifts dramatically, as highlighted by the yellow boxes in Fig.\,\ref{FIG3_NHQ}\,(a). Previous studies have revealed that min-mode methods, such as GAD, fail to evolve along the MEP on this PES~\cite{henkelman1999dimer,zhang2012shrinking}, making it an ideal test case for comparing the performance of SPM with min-mode methods.

Fig.\,\ref{FIG3_NHQ}\,(b-d) illustrates the evolution paths (red curves) of the dimer method, GAD, and SPM on the PES of $V_2(\mathbf{x}_1,\mathbf{x}_2)$, starting from different initial positions (I1, I2, I3, I4). Although all three methods successfully converge to the saddle point S1, the evolution paths  of the dimer method (b) and GAD (c) deviate more significantly from the MEP compared to SPM (d).
Moreover, the evolution paths of the dimer method and GAD show a strong dependence on the initial positions, resulting in drastically different trajectories. In contrast, the SPM evolution paths remain nearly identical, regardless of the starting point, showcasing its robustness and consistency.
Specifically, the points with the highest energy along the evolution paths of the dimer method and GAD (green dots in Fig.\,\ref{FIG3_NHQ}\,(b,c)) do not always coincide with the saddle point S1, further highlighting the significant differences between evolution paths and the MEP. Conversely, the highest energy point along the SPM evolution paths (green dot in Fig.\,\ref{FIG3_NHQ}\,(d)) consistently aligns with the saddle point S1, demonstrating SPM's  ability to closely follow the MEP, even in the presence of complex energy landscapes.

Fig.\,\ref{FIG3_NHQ}\,(e) visualizes the details of SPM evolution starting from the initial position I3, which is in the opposite direction of the MEP. Even when starting from the opposite direction of the MEP, the drifting dynamics of SPM effectively reorients the spring pair and pulls it back onto the MEP. Although the SPM trajectory may slightly deviate from the MEP at the sharp turns, it quickly backs to the MEP after navigating through these regions, demonstrating the robustness of using the MEP as a guiding compass.

In summary, these numerical experiments on the simple two-dimensional test functions clearly demonstrate the fundamental differences between SPM and min-mode methods in their approach to finding saddle points. While the dimer method and GAD rely on local curvature information and may deviate significantly from the MEP, SPM consistently utilizes the MEP as a global guide, ensuring more efficient and reliable convergence to saddle points.

% The illustrative examples on the simple two-dimensional functions have elucidated the operating mechanism of SPM and its differences from min-mode methods. In the next section, we apply SPM to explore complex energy landscapes in real physicochemical systems, showcasing its potential for handle high-dimensional problems.

\subsection{Lennard-Jones clusters}
In this section, we apply SPM to study the ground states and transition states of Lennard-Jones (LJ) clusters, which provide a realistic simulation of micro-particle interactions. The LJ potential between particle pair is
\begin{equation}
v_{\mathrm{LJ}}(r)=\frac{1}{r^{12}}-\frac{2}{r^{6}},
\end{equation}
where $r$ represents the distance between two particles. The objective is to find all local minima and saddle points on the high-dimensional PES $V_{\mathrm{LJ}}(\mathbf{r}^{N})$,
\begin{equation}
V_{\mathrm{LJ}}(\mathbf{r}^{N})=\sum_{i=1}^{N-1}\sum_{j=i+1}^{N}v_{\mathrm{LJ}}(|\mathbf{r}_{i}-\mathbf{r}_{j}|),
\end{equation}
where $\mathbf{r}^{N} = (\mathbf{r}_{1},\mathbf{r}_{2},...,\mathbf{r}_{N})$ represents the positions of all $N$ particles.

We focus on a system of 7 particle Lennard-Jones clusters (LJ7) and find all the minima and the lowest energy saddle points connecting different minima. 
Previous studies~\cite{Davis1990Exploring,Tsai1993Use,Miller1997Isomerization} have shown that the LJ7 system has 4 local minima and 12 saddle points, with 6 saddle points connecting different minima and 6 saddle points connecting identical minima.
The 5 lowest energy saddle points connecting different minima are particularly significant, as they reveal the most likely transition pathways between stable configurations.

\begin{figure}
        \centering	
		\includegraphics[width=0.9\columnwidth]{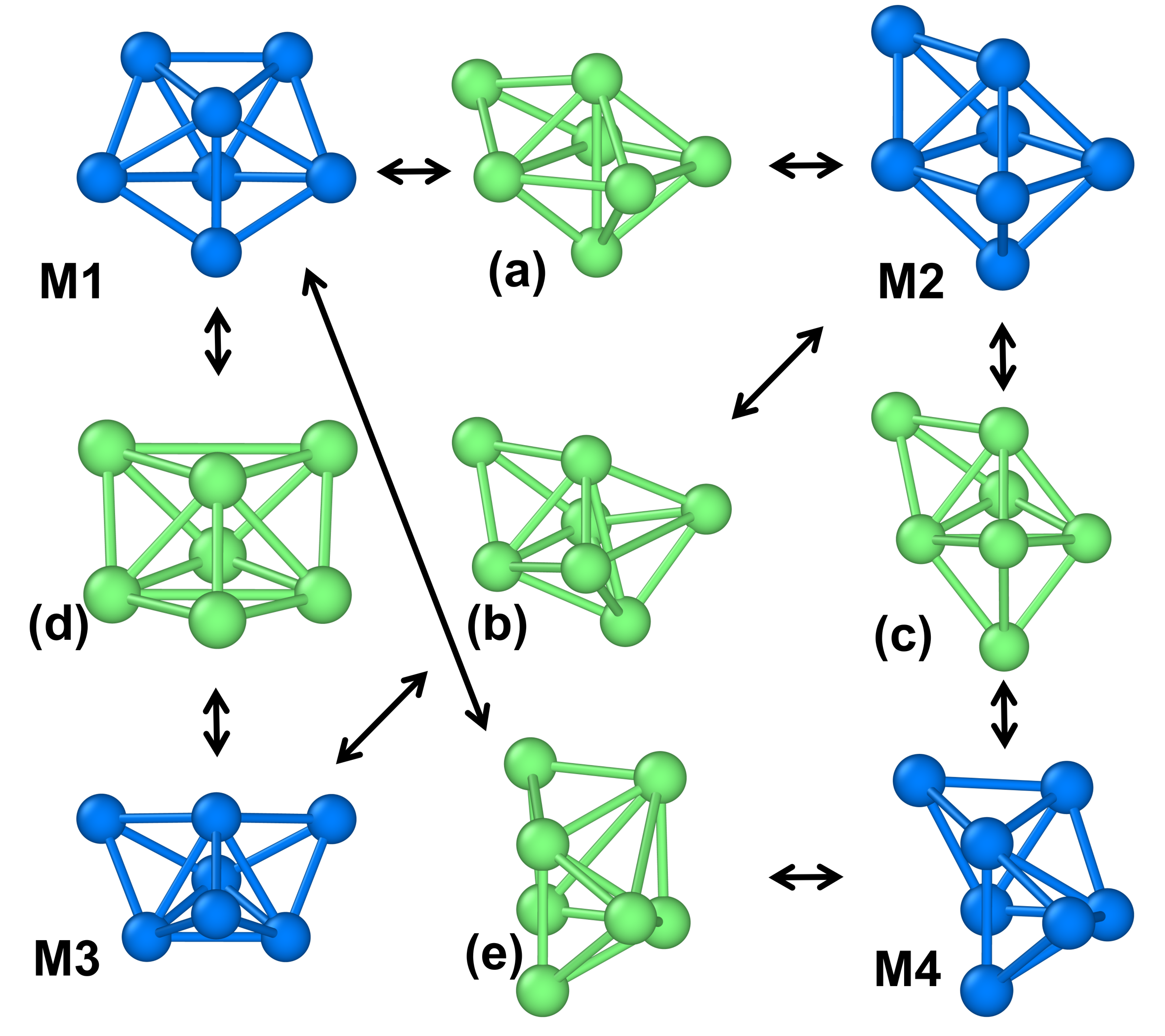}
        \caption{Conformations of minima and saddle points and the connections between them. The minima M1-M4 are shown in blue, the saddle points (a-e) in green, and the black arrows indicate possible transition paths between them. The results are consistent with previous research~\cite{Miller1997Isomerization}.
        }\label{LJ}
\end{figure}

Specifically, starting from one known minimum with random perturbations, we employ SPM to locate the connected saddle points. By performing gradient descent along the obtained unstable mode from each saddle point, we can reach another local minimum. This process is repeated for each newly found minimum until all 5 lowest-energy saddle points (a-e) connecting different minima and their associated minima (M1-M4) are identified, as illustrated in Fig.\,\ref{LJ}.

The energies of four minima are listed in Table \ref{tab:min-ene-LJ}, while the energies and energy barriers of the saddle points between minima are provided in Table \ref{tab:saddle-ene-LJ}. The transition paths (Fig.\,\ref{LJ}) and energy barriers (Table \ref{tab:saddle-ene-LJ}) can help predict the most probable path for cluster evolution. As an illustrative example, let us examine a transition from the local minimum M4 to the global minimum M1 in Fig.\,\ref{LJ}. There are two possible routes: a one-step transition M4 $\rightarrow$ M1, and a two-step transition M4 $\rightarrow$ M2 $\rightarrow$ M1. The one-step transition has a higher energy barrier of $\Delta E = 0.5067$, corresponding to saddle point (e). In contrast, the two-step transition has two lower energy barriers, $\Delta E = 0.2497$ and $\Delta E = 0.4903$, associated with saddle points (c) and (a), respectively. Therefore, the two-step transition from M4 to M1 via M2 is more likely to occur due to its lower overall energy barrier compared to the one-step transition. SPM requires minimal prior knowledge of the system, needing only one known minimum as a starting point. SPM can efficiently discover a large number of other minima and their connecting saddle points, enabling the exploration of global minima in large LJ systems and the study of phase transition mechanisms in complex particle clusters.
\renewcommand{\arraystretch}{1.3}
\begin{table}
\fontsize{8}{9.6}\selectfont
\caption{Energies of LJ7 cluster minima. The values for the minima reported here have been obtained by gradient descent from the saddle points along the unstable mode.}
\begin{tabular*}{\columnwidth}{@{\extracolsep{\fill}}ccc}
\hline
\hline
Minima & Configuration & Energy \\
\hline
M1 & Pentagonal bipyramid & -16.5054 \\
M2 & Capped octahedron & -15.9350 \\
M3 & Capped octahedron & -15.5932 \\
M4 & Capped octahedron & -15.5932 \\
\hline
\end{tabular*}
\label{tab:min-ene-LJ}
\end{table}

\begin{table}
\fontsize{8}{9.6}\selectfont
\caption{Energies and energy barriers of LJ7 cluster saddle points obtained by SPM.}
\begin{tabular*}{\columnwidth}{@{\extracolsep{\fill}}cccc}
\hline
\hline
Saddle Points & Connected Minima & Energy & Energy Barrier \\
\hline
(a) & M2-M1 & -15.4447 & 0.4903 \\
(b) & M3-M2 & -15.3199 & 0.2733 \\
(c) & M4-M2 & -15.2834 & 0.2497 \\
(d) & M3-M1 & -15.0334 & 0.5595 \\
(e) & M4-M1 & -15.0264 & 0.5067 \\
\hline
\end{tabular*}
\label{tab:saddle-ene-LJ}
\end{table}

% {\it Lifshitz-Petrich energy functional.--}
\subsection{Lifshitz-Petrich energy functional}
The LP model is an effective theory that can describe quasiperiodic structures like soft matter quasicrystals~\cite{lifshitz1997theoretical,jiang2015stability,lifshitz2007soft}.
In particular, the LP free energy is 
\begin{eqnarray}\label{eq:LP}
	E_{LP}(\mathbf{\phi}) &=& \frac{1}{| \Omega |}\int_{\Omega}   \frac{1}{2}[(q_1^2 + \Delta)(q_2^2 + \Delta)\mathbf{\phi}]^2 \nonumber\\
	&-&\frac{\varepsilon}{2}\mathbf{\phi}^2 - \frac{\alpha}{3}\mathbf{\phi}^3 + \frac{1}{4} \mathbf{\phi}^4   \, d\mathbf{r}, 
\end{eqnarray}
where $\mathbf{\phi}(\mathbf{r})$ is a real-valued function that measures the order of system. $\Omega$ is a bounded domain of the system with the volume $|\Omega|$. $\varepsilon$ and $\alpha$ are coefficients of measuring the temperature and the intensity of three-body interaction, respectively. $q_1$ and $q_2$ are two characteristic length scales, which are necessary to stabilize quasicrystals.

\begin{figure*}[!hbpt]
        \centering	
		\includegraphics[width=1.0\columnwidth]{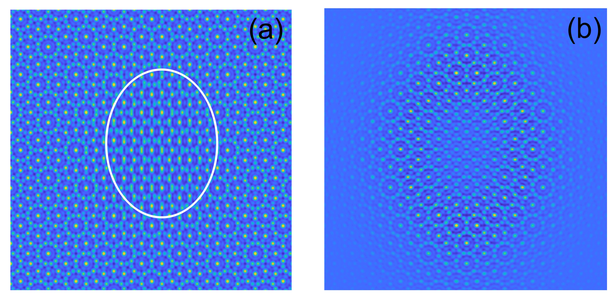}
        \includegraphics[width=1.0\columnwidth]{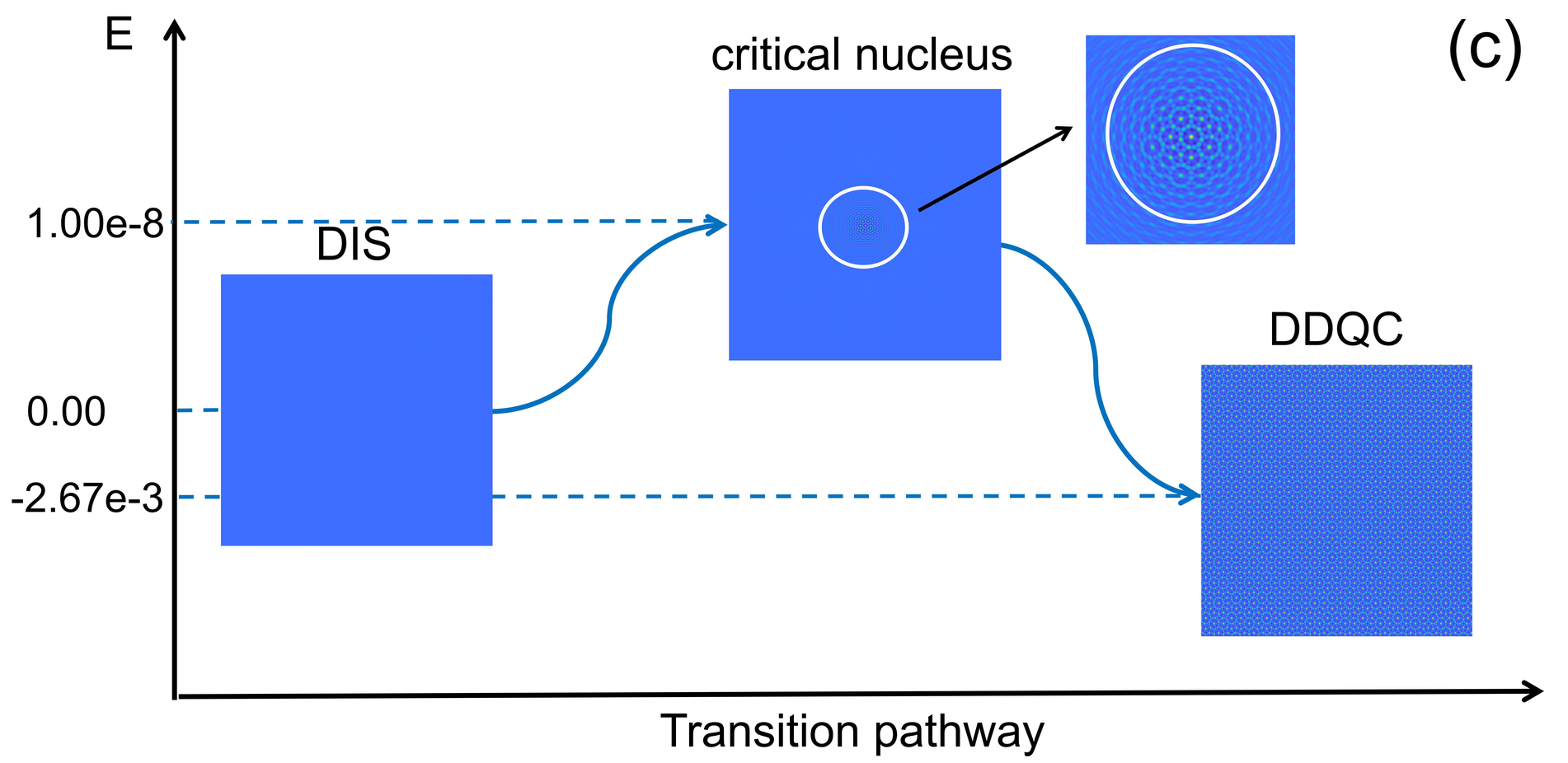}\\
        \caption{Locating the critical nucleus and MEP for the DIS-DDQC phase transition using SPM. (a) The critical nucleus configuration exhibits dodecagonal symmetry (white circle). (b) The unstable mode at the  saddle points is represented by the spring direction. (c) Schematic MEP between disorder and DDQC obtained by gradient descent along the unstable mode.
          }\label{FIG3}
\end{figure*} 

In this article, we set $q_1 = 1$, $q_2 = 2\cos(\pi/12)$ to stabilise the dodecagonal quasicrystal (DDQC)~\cite{lifshitz1997theoretical}.
Furthermore, we impose the following mean zero condition $(1/|\Omega|)\int_{\Omega} \mathbf{\phi}(\mathbf{r})\,d\mathbf{r}  = 0$ of order parameter on the LP systems to ensure the mass conservation.

We employ the Fourier pseudo-spectral method to discretize the LP energy functional as done in ~\cite{yin2021transition}. The considered domain $\Omega$ is discreted as a grid system, with $N$ nodes in each dimenstion. The function $\mathbf{\phi}(\mathbf{r})$ is represented by the values at these nodes, yielding a $N^2$-dimensional variable set.
The initial spring pair $\mathbf{\phi}^{0}_1$ is placed near a local minimum, with $\mathbf{\phi}^{0}_2 = \mathbf{\phi}^{0}_1 + \xi  \mathbf{v}^{*}$, where $\xi$ is a small threshold and $\mathbf{v}^{*}$ is a random perturbation direction.
In fact, if we know some empirical perturbation directions related to the final state, it will speed up the saddle points calculation tremendously.

\begin{figure*}[!hbpt]
        \centering	
		\includegraphics[width=1.0\columnwidth]{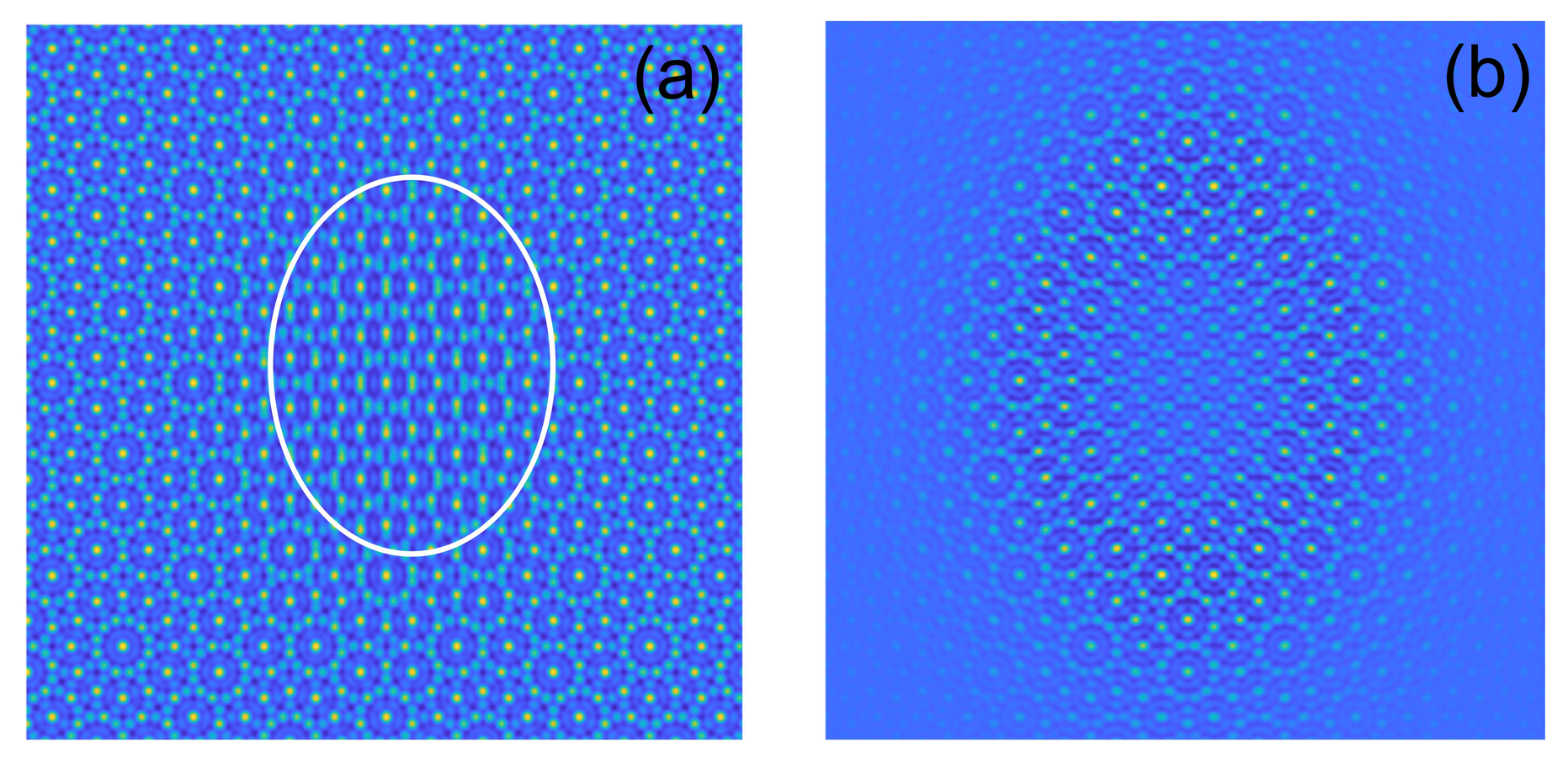}
        \includegraphics[width=1.0\columnwidth,trim={0 0.01cm 0 0},clip]{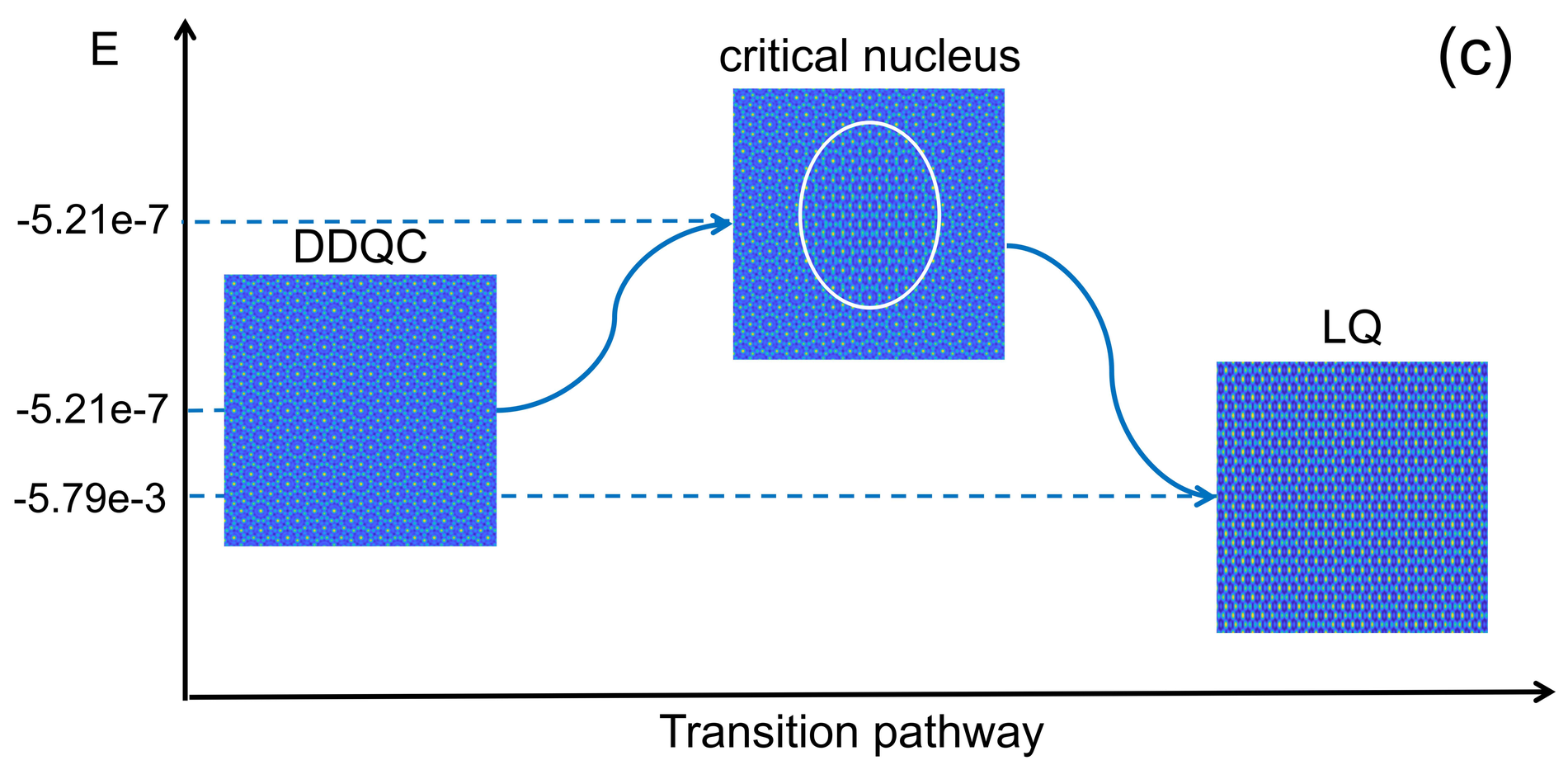}\\
        \caption{Locating the critical nucleus and MEP for the DDQC-LQ phase transition using SPM. (a) The critical nucleus configuration shows a layered structure (white circle). (b) The unstable mode at the  saddle points is represented by the spring direction. (c) Schematic MEP between DDQC and LQ obtained by gradient descent along the unstable mode.
          }\label{FIG4}
\end{figure*}

Then we apply SPM to locate transition states in the LP model for two cases: (1) nucleation from Disordered (DIS) to DDQC; (2) phase transition from DDQC to Lamellar quasicrystal (LQ). Case 2 involves escaping a degenerate minimum corresponding to DDQC.
% For both examples, the spring parameters are: $l_s=10^{-3}, \alpha_1 =2, \alpha_2 = 10^{-3}, \alpha_3 = 2$, stopping criterion for drifting is $\epsilon_1=10^{-5}$, the convergence criterion is $\epsilon_2 = 10^{-7}$ , and maximum iteration in drifting process is $n^{(d)} = 200$. 

For the DIS to DDQC phase transition, we set $\Omega = [0,60\pi)^2, N = 512, \epsilon = -0.01,  \alpha = 1.0$. The critical nucleus and the unstable mode are accurately calculated by SPM. The critical nucleus configuration shows the characteristic dodecagonal symmetry (Fig.\,\ref{FIG3}\,(a)). The unstable mode (Fig.\,\ref{FIG3}\,(b)) resembles the critical nucleus, indicating the direction of DDQC growth and decay. By slightly perturbing along this mode and using gradient descent, the MEP connecting disorder and DDQC is obtained, as shown in Fig.\,\ref{FIG3}\,(c).

For phase  transition of DDQC to LQ phase transition, we set $\Omega = [0,224\pi)^2, N = 1024, \epsilon = 0.05,  \alpha = 1.0$. The critical nucleus (Fig.\,\ref{FIG4}\,(a)) exhibits LQ order in the center of DDQC. In the unstable mode (Fig.\,\ref{FIG4}\,(b)), the interface between the critical nucleus and DDQC shows a tendency to transform into LQ.
DDQC is a degenerate minimum whose Hessian has four zero  eigenvalues~\cite{jiang2015stability,cui2023efficient}. Conventional min-mode methods may struggle to escape such degenerate minima, as they rely on the eigenvector of the smallest eigenvalue of Hessian as the climbing direction~\cite{yin2021transition}. In contrast, SPM uses the spring orientation aligned with the MEP tangent as a natural climbing direction, which enables it to efficiently escape such degenerate minima and locate the  saddle points.
Again, performing gradient descent optimization from the critical nucleus along the unstable mode, the MEP connecting DDQC and LQ is obtained, as shown in Fig.\,\ref{FIG4}\,(c). The MEP provides the evolutionary process of how the LQ emerges and grows within the DDQC.

\section{Summary}
In this work, we propose a novel SPM for efficiently locating saddle points on complex high-dimensional PES without requiring Hessian information. By evolving a simple spring-coupled particle pair under tailored gradient-based dynamics, SPM naturally acquires the MEP tangent and converges to the saddle points.

Compared with the min-mode methods that rely on the local lowest curvature mode, SPM leverages the global perspective of the MEP to guide the search dynamics. This enables SPM to more reliably find the saddle points connected to the initial state, making it less sensitive to perturbations applied to the starting points. Furthermore, SPM significantly reduces the computational cost by dealing with only one spring pair, compared to path-finding methods that evolve the entire path.

The efficiency and applicability of SPM have been demonstrated through a series of examples, ranging from two-dimensional test functions to physical systems, including the determination of transition states in a 7-particle LJ cluster and the identification of critical nuclei in quasicrystal formation described by LP free energy functional.

% In summary, we have presented a novel spring pair method for efficiently locating  saddle points on complex high-dimensional PES without requiring Hessian information. By evolving a simple spring-coupled particle pair under tailored gradient-based dynamics, the technique naturally acquires the MEP tangent and converges to the saddle points.
% We have demonstrated the implementation and performance of SPM on a two-dimensional function and the LP energy functional involving quasicrystal phase transition. The SPM precisely locates the critical nuclei and unstable mode of quasicrystals phase transition. In particular, it readily handles the translationally invariant degenerate minima, which presents challenges for conventional eigenvector-direction-based techniques.
% Overall, we develop a straightforward yet powerful tool for exploring complex PES and locating transition states of phase transitions.
% While showing considerable promise, future work should focus on expanding the efficiency and applicability of SPM, including investigating diverse complex systems, incorporating advanced optimization protocols.

\section*{Data Availability Statement}
The data that support the findings of this study are available
from the corresponding authors upon reasonable request.

\bibliography{aipsamp}% Produces the bibliography via BibTeX.

\end{document}